\documentclass[12pt]{article}

\usepackage{amsfonts,amssymb,amsmath, amsthm, color, a4wide}
\usepackage{mathtools}
\usepackage{mathrsfs}
\usepackage{hyperref}
\usepackage{graphicx}
\usepackage{epstopdf}
\usepackage{bbm}
\usepackage{multirow}

\usepackage[round]{natbib}
\setcitestyle{notesep={: }}
\setcitestyle{aysep={,}}

\usepackage{authblk}

\begin{document}
 
\title{%Likelihood based inference for 
Should we condition on the number of points when modelling spatial point patterns?} 
\author[1]{Jesper M\o ller}
\author[1]{Ninna Vihrs} 
\affil[1]{Department of Mathematical Sciences,
 Aalborg University, Skjernsvej 4A, DK-9220 Aalborg, email:
\texttt{jm@math.aau.dk,nvihrs@math.aau.dk}}

%\date{\date{}}
\maketitle

\begin{abstract}
We discuss the practice of directly or indirectly assuming a model for the number of points when modelling spatial point patterns even though it is rarely possible to validate such a model in practice since most point pattern data consist of only one pattern. We therefore explore the possibility to condition on the number of points instead when fitting and validating spatial point process models. In a simulation study with different popular spatial point process models, we consider model validation using global envelope tests based on functional summary statistics. We find that conditioning on the number of points will for some functional summary statistics lead to more narrow envelopes and thus stronger tests and that it can also be useful for correcting for some conservativeness in the tests when testing composite hypothesis. However, for other functional summary statistics, it makes little or no difference to condition on the number of points. When estimating parameters in popular spatial point process models, we conclude that for mathematical and computational reasons it is convenient to assume a distribution for the number of points. 
\end{abstract}

\noindent\textit{Keywords:} 
conditional inference; global envelope test; maximum likelihood; maximum pseudo-likelihood; 
spatial point process models.

%\begin{document}
%
%\maketitle
%
%\begin{abstract}
%
%\end{abstract}

%\keywords{conditional inference; global envelope test; maximum likelihood; maximum pseudo-likelihood; spatial point process models}

\section{Introduction}
\label{s:1.1}
Consider a spatial point process defined on the $d$-dimensional Euclidean space $\mathbb R^d$, and let $W\subset\mathbb R^d$ be a bounded region $W\subset\mathbb R^d$ within which realizations of the process are observed. In the literature on statistical analysis of spatial point processes, the data usually consist of a single realization $\{x_1,\ldots,x_n\}\subset W$, that is, a (finite) point pattern where both the number of points $n$ and the locations $x_1,\ldots,x_n$ are considered to be random. For instance in the \texttt{R}-package \texttt{spatstat} \citep{spatstat}, which is widely used for analysing spatial point patterns, there are more than fifty data examples of point patterns, but only seven of these consist of more than one point pattern. Although it is impossible from a single point pattern to validate the plausibility of any claimed model for the number of points, the practice is nonetheless to fit, validate and use spatial point process models which directly or indirectly assume a model for the number of points. 

For the example of a stationary Poisson process, $n$ is a realization of a Poisson distributed random variable, and conditioned on $n$, the points are realizations of $n$ independent uniformly distributed random variables on $W$. Even for this simple example of a point process model, it is not possible to validate the assumed Poisson distribution for the number of points based on just one realization. A common procedure for testing whether a Poisson process model fits a given point pattern is to divide $W$ into disjoint subsets of equal size and count the number of points falling within each subset. Conditioning on $n$, these observed counts constitute a realization from a multinomial model with equal probabilities, and the validity of this model can be checked using a goodness-of-fit test, e.g.\ as implemented in the function {\bf{\sf quadrat.test}} from \texttt{spatstat}. However, even if we accept this multinomial model, to claim that $n$ is a realization from a Poisson distribution amounts to assume independence for the counts \citep{Moran}, and indeed this assumption is hard to justify by a statistical test. Therefore, it will be impossible from a single point pattern to validate that $n$ is a realization from a Poisson distribution.

We still believe that it is usually reasonable to think of the number of points as a single realization of a random variable. We merely point out that it would be inappropriate to make claims about the distribution of the number of points since we have no chance to validate these as illustrated for a stationary  Poisson process above. If we are indeed willing to make some model assumptions and can make simulated point patterns under these assumptions, such simulations can of course be used to estimate the distribution of the number of points under the model if wanted, but why should we believe in such a distribution when we do not have the means to validate this? It may thus be more fair to accept that we do not know anything about the distribution of the number of points and therefore use a conditional approach instead. However, we have not found much places in the spatial point process literature where the role of conditioning on $n$ is studied. \cite{Ripley:77} provided a short discussion of conditional inference saying `For Poisson models we can justify this conditioning by an appeal to a conditionality principle. In general all we can say is that conditional inference seems reasonable and that our revised procedures yield valid conditional tests'. Further, \cite{Ripley:88} wrote `Where we are interested in interactions, $n$ may be approximately ancilliary'. He illustrated this for a Strauss process which on the unit square when ignoring boundary effects has density
\[f(\{x_1,\ldots,x_n\})\propto \beta^n\gamma^{s(\{x_1,\ldots,x_n\})}\]
where $\beta>0$, $0\le\gamma\le 1$, $R>0$ and $s(\{x_1,\ldots,x_n\})$ is the number of $R$-close pairs of points in the point pattern; we return to this process in Sections~\ref{s:models_sim} and \ref{s:Gibbs_est} and Appendix~\ref{app:MLEGibbs}.  \cite{Ripley:88} derived an approximation of the probability density function for the number of points:
\begin{equation}\label{e:ripley}
p(n)\propto \frac{\beta^n}{n!}\exp\{(\gamma-1)n(n-1)\pi R^2/2\}.
\end{equation}
He noticed that this density depends on $\gamma$ but wrote `the dependence is quite weak in typical cases' where he referred to a plot of the cumulative distribution functions when $\beta=50$, $R=0.05$ and $\gamma=0.0,0.2,0.4,0.8,1.0$. Mean values of $n$ in this case and when $\beta=200$ are shown in Table~\ref{tab:mean_n} where we see a good agreement between those means calculated from the approximation in \eqref{e:ripley} and those obtained by simulations. In our opinion, the mean values in Table~\ref{tab:mean_n} depend much on $\gamma$, but we notice in Section~\ref{s:Gibbs_est} that maximum pseudo-likelihood estimates do not depend much on whether or not we condition on $n$. 

\begin{table}[ht!]
\centering
\begin{tabular}{l|cccccc}
 & $\gamma = 0$ & $\gamma = 0.2$ & $\gamma = 0.4$ & $\gamma = 0.6$ & $\gamma = 0.8$ & $\gamma = 1$\\ \hline
\multirow{2}{*}{$\beta = 50$} & 37.33 & 39.13 & 41.19 & 43.61 & 46.48 & 50.00\\
 & (36.91) & (38.91) & (41.24) & (43.78) & (46.56) & (49.87)\\
\multirow{2}{*}{$\beta = 200$} & 95.00 & 104.11 & 115.92 & 132.12 & 156.45 & 200.00\\
 & (88.37) & (100.72) & (114.53) & (132.35) & (157.36) & (200.32)
\end{tabular}
\caption{Mean values for the number of points in a Strauss process on a two-dimensional unit square with $R=0.05$ and different values of $\gamma$ and $\beta$. The means are calculated from the approximate distribution in \eqref{e:ripley}, and the numbers in parenthesis are the means obtained from 5000 simulations of the process}
\label{tab:mean_n}
\end{table}

Apart from Ripley's study of the Strauss process considered above we are not aware of any thorough study of the effect of conditioning on $n$.  Moreover,
the practice is still to work with spatial point process models without conditioning on $n$. Therefore, the objective of this paper is to investigate the consequences of this practice and to explore the possible benefits of conditioning on $n$ when considering various popular classes of spatial point process models. We investigate this through a comprehensive simulation study of model validation based on the widely used method of global envelopes and corresponding tests \citep{GET2017}, and by discussing the effect of conditioning on $n$ when making frequentistic parameter estimation.

Our paper is organized such that Section~\ref{s: background} contains some preliminaries needed for our main contributions in Sections~\ref{s:simstudy_GET}--\ref{s:conclusion}, where Section~\ref{s:conclusion} summarises our findings. Technical details related to these sections are found in the appendices. All statistical analyses were made with \texttt{R} \citep{R}. We used the packages \texttt{spatstat} \citep{spatstat} for handling spatial point patterns, \texttt{GET} \citep{GETinR} to make global envelope tests and \texttt{ggplot2} \citep{ggplot} for visualisation. Furthermore, we used our own implementations of simulation and estimation procedures when conditioning on the number of points, which can be found in the \texttt{R}-scripts in the ancillary files. 

\section{Preliminaries}\label{s: background}

\subsection{Setting and notation}\label{s:set}

Throughout this paper we use the following point process setting and notation.

For a subset $x\subset\mathbb R^d$, let $n(x)$ denote its cardinality (setting $n(x)=\infty$ if $x$ is not a finite set) and $x_B\coloneqq x\cap B$ its restriction to any set $B\subseteq\mathbb R^d$. Let $\Omega$ denote the set of all locally finite subsets $x\subset\mathbb R^d$, that is, $n(x_B)<\infty$ whenever $B\subset\mathbb R^d$ is bounded. A simple locally finite point process on $\mathbb R^d$ is a random variable taking values in $\Omega$. Here and elsewhere, we omit measure theoretical details and refer instead to \cite{textbook} and the references therein. The point process can also be specified in terms of the counting measure $N(B)\coloneqq n(X_B)$ for bounded sets $B\subseteq\mathbb R^d$ (more precisely, $B$ should also be a Borel set, but as mentioned we omit such details). 

We assume that $X$ is a stationary simple locally finite point process on $\mathbb R^d$ with intensity $\rho\in(0,\infty)$. This means that $X\in\Omega$, the distribution of $X$ is invariant under translations in $\mathbb R^d$ and $\mathrm EN(B)=\rho|B|$ where $|B|$ denotes the $d$-dimensional volume (Lebesgue measure) of $B\subseteq\mathbb R^d$. Stationarity is a common assumption for spatial point processes, and it allows us to deal with frequently used functional summary statistics, see Section~\ref{s:summary-stat}. We also assume that a single realization $x=\{x_1,\ldots,x_n\}$ of $X_W$ has been observed where the observation window $W\subset\mathbb R^d$ is compact and $|W|>0$.

Let $u,v\in\mathbb R^d$, $r\ge0$ and $B\subset\mathbb R^d$. Then $\mathbbm{1}(\cdot)$ is the indicator function; $\|u-v\|$ is the usual distance between $u$ and $v$; $b(u,r)$ is the closed $d$-dimensional ball with centre $u$ and radius $r$; ${\mathrm{dist}}(B,u)\coloneqq\inf\{r>0\,|\, b(u,r)\cap B\not=\emptyset\}$ is the distance from $u$ to $B$; $B_{\ominus r}\coloneqq\{u\in\mathbb R^d\,|\,b(u,r)\subset B\}$ is $B$ eroded by a ball of radius $r$; $B_{\oplus}\coloneqq\cup_{u\in B}b(u,r)$ is $B$ dilated by a ball of radius $r$; and $(B_{\ominus r})_{\oplus r}\subseteq B$ is the opening of $B$ by a ball of radius $r$. Finally, we use the convention $0/0\coloneqq 0$. 

\subsection{Functional summary statistics}\label{s:summary-stat}

Functional summary statistics $\hat K,\hat F,\hat G\text{ and }\hat J$ which are non-parametric (empirical) estimates of the theoretical functions below are widely used for exploratory purposes, model fitting and model checking, see \cite{spatstat} and the references therein. 

For every $r>0$ and an arbitrary point $u\in\mathbb R^d$, Ripley's $K$-function is defined by
\begin{equation}\label{eq:K}
\rho K(r)\coloneqq \mathrm E(N(b(u,r)\setminus\{u\})\mid u\in X),
\end{equation}
the empty space function $F$ (or spherical contact function) and the nearest-neighbour function $G$ are
\begin{equation}\label{eq:FG}
F(r)\coloneqq \mathrm P({\mathrm{dist}}(X,u)\le r),\qquad
G(r)\coloneqq \mathrm P({\mathrm{dist}}(X\setminus\{u\},u)\le r\mid u\in X)
\end{equation}
and for $F(r)<1$ the $J$-function is
\begin{equation}\label{eq:J}
J(r)\coloneqq (1-G(r))/(1-F(r)).
\end{equation}
These definitions do not depend on the choice of $u$ since $X$ is stationary, and as indicated by the notation we have conditioned on $u\in X$ in the definitions of $K$ and $G$, meaning that $X\setminus\{u\}$ then follows the reduced Palm distribution of $X$ at $u$. 

When estimating the $K,F,G\text{ and }J$-functions by non-parametric methods, different types of edge correction methods have been suggested in order to adjust for the fact that $n(x_{W\cap b(u,r)})$ tends to be smaller for points $u\in W$ which are close to the boundary of $W$ compared to points $u\in W$ which are far from the boundary of $W$. \cite{spatstat} noticed that the choice of edge correction method is usually not very important. We choose to use a particular border (or minus-sampling) correction method which is available in \texttt{spatstat} and refer to \citet{spatstat} for the concrete estimates. 

When conditioning on $n$, we are not aware of how to modify the definitions of $K$, $F$, $G$ and $J$. If we just condition in \eqref{eq:K}--\eqref{eq:J}, the expressions will depend on $u$ because stationarity no longer holds. Still, when only a single point pattern is observed and hence it is impossible to validate any claimed model of $N(W)$, it seems appropriate to condition on $n$ when calculating global envelopes and tests as in Section~\ref{s:simstudy_GET}.

\subsection{Models} 
\label{s:models_sim}

For the simulation studies in Sections~\ref{s:simstudy_GET} and \ref{s:Gibbs_est}, we consider four concrete examples of point process models on $\mathbb{R}^2$ as specified in M1--M4 below. Appendix~\ref{app:models} provides further details including how to make simulations both with and without conditioning on the number of points. 

M1: $X$ is a stationary Poisson process. This is the model of no spatial interaction or complete spatial randomness. When making simulations in Section~\ref{s:simstudy_GET}, we let the intensity be $\rho=100$.

M2: $X$ is a stationary log-Gaussian Cox process. Then, $X$ is driven by a stochastic intensity $Z = \exp(Y)$ where $Y$ is a stationary Gaussian random field on $\mathbb R^2$, meaning that $X$ conditioned on $Y$ is a Poisson process with intensity function $Z$ \citep{LGCP}. We use an exponential covariance function $c(u,v) \coloneqq \sigma^2\exp \left (-\|u - v\|/\delta\right )$ for $u,v\in\mathbb R^2$ where $\sigma^2>0$ is a variance parameter and $\delta>0$ is a scale parameter, and so the mean of $Y$ is $\mu=\log(\rho)-\sigma^2/2$. Because of the positive correlation in $Y$, realizations of $X$ exhibit clustered behaviour. When making simulations in Section~\ref{s:simstudy_GET}, we let $\rho=100$, $\sigma^2=1$ and $\delta=0.1$. 

M3: $X$ is a stationary Strauss process. This process has parameters $\beta>0$, $0\le\gamma\le1$ and $R>0$. It is defined by a so-called local specification: Let $B\subset\mathbb R^2$ be a bounded set of positive area and define the $R$-close neighbourhood to $B$ by $\partial B\coloneqq B_{\oplus R}\setminus B$. Then, for every such $B$, $X_B$ and $X_{\mathbb{R}^2\setminus B_{\oplus R}}$ are conditionally independent given $X_{\partial B}$. Furthermore, for every finite $x_{\partial B}\subseteq \partial B$, $X_B$ conditioned on $X_{\partial B}=x_{\partial B}$ has a density with respect to a Poisson process of intensity 1 and restricted to $B$. This conditional density is 
\begin{equation}\label{e:cond-strauss-dens}
f_{B}(x_B\,|\,x_{\partial B})\propto \beta^{n(x_B)}\gamma^{s(x)} 
\end{equation}
for finite $x_B\subseteq B$, $x=x_B\cup x_{\partial B}$ and $s(x)=\sum_{i<j}\mathbbm 1(\|x_i-x_j\|\le R)$ if $x=\{x_1,\ldots,x_n\}$. The normalizing constant which is omitted in \eqref{e:cond-strauss-dens} depends on $(\beta,\gamma,R)$ and $x_{\partial B}$, and it is intractable unless $\gamma=1$. When $\gamma=1$, $X$ is just a stationary Poisson process with intensity $\beta$. As $\gamma$ decreases, $X$ becomes more and more inhibitory, and it is a Gibbs hard core model if $\gamma=0$. When making simulations in Section~\ref{s:simstudy_GET}, we let $\beta= 200$, $\gamma = 0.3$ and $R = 0.05$. Then, the intensity is approximately 100.

M4: $X$ is a stationary determinantal point process. In brief, determinantal point processes \citep{Macchi,DPP} are specified by a function $C:\mathbb R^d\times\mathbb R^d\mapsto \mathbb C$ called the kernel, and they are repulsive at all scales, cf.\ Appendix~\ref{app:DPP_def_sim}. We use a Gaussian kernel $C(u,v)=\rho\exp(-\|(u-v)/\kappa\|^2)$ where $\kappa>0$ is a scale parameter and $\rho>0$ is the intensity. It should be satisfied that $\kappa\leq 1/\sqrt{\rho\pi}$, and this upper limit corresponds to the most repulsive case when $\rho$ is fixed. When making simulations in Section~\ref{s:simstudy_GET}, we let $\rho = 100$ and $\kappa = 0.03 \leq 1/\sqrt{100\pi}\approx0.056$.

\section{Global envelopes and the effect of conditioning}
\label{s:simstudy_GET}
In this section, we investigate the effect of conditioning on the number of points when using global envelopes for model validation. Section~\ref{s:setup} first describes the set-up of the simulation study, and Section~\ref{s:results} describes and interpret the results.

\subsection{Set-up}\label{s:setup}
We investigate the effect of conditioning on the number of points when considering global envelopes for functional summary statistics and corresponding tests based on the extreme rank length as described in \citet{GET2017}, \citet{GET2018} and \citet{GETinR}. Briefly, a $(100-\alpha)$\% global envelope consists of a lower and an upper curve defining a region such that the observed functional summary statistic for data falls completely between these bounding curves if and only if the global envelope test cannot be rejected at level $\alpha$\%. There of course exist other tests which can be used in connection with spatial point processes, see e.g.\ \citet[chapter 10]{spatstat}; however, since the use of a global envelope and its corresponding test statistic is by far the most popular method for performing model validation of a fitted spatial point process model, we restrict attention to this test procedure. There are ways to make a combined global envelope test based on several functional summary statistics, but for our purpose we prefer to investigate the effect of conditioning on the number of points for each functional summary statistic. 

We made the simulation study as follows. Under each of the four models M1--M4 we simulated 1000 independent point patterns within a two-dimensional unit square (the observation window $W$). For each of these point patterns, we fitted the parameters of the models as described in the last paragraph of this section. Under each fitted model and each true model, we made further 2500 simulations with and 2500 simulations without conditioning on the number of points. From each of these four cases, or three in the case of the Poisson process since the fitted and true model is the same when conditioning on the number of points, we used the 2500 simulations to calculate 95\% global envelopes based on each of the functional summary statistics $\hat F, \hat G, \hat J$ and $\hat K$. Some further technical and practical details related to the set-up of the simulation study are deferred to Appendix~\ref{app:simdetails}. 

Clearly more narrow envelopes are preferable when comparing envelopes for the same type of functional summary statistic. For simplicity, in order to spot a general tendency in the width of envelopes, we considered for each envelope a numerical approximation of its area $\int_{0}^{R}(c_u(r) - c_l(r))\,\text{d}r$ where $c_u$ and $c_l$ are the upper and lower curves of the envelope, respectively, and $R$ is the highest $r$-value for which the considered functional summary statistic was estimated.

\subsection{Results}\label{s:results}
\begin{figure}[htp!]
\centering
\includegraphics[scale=1]{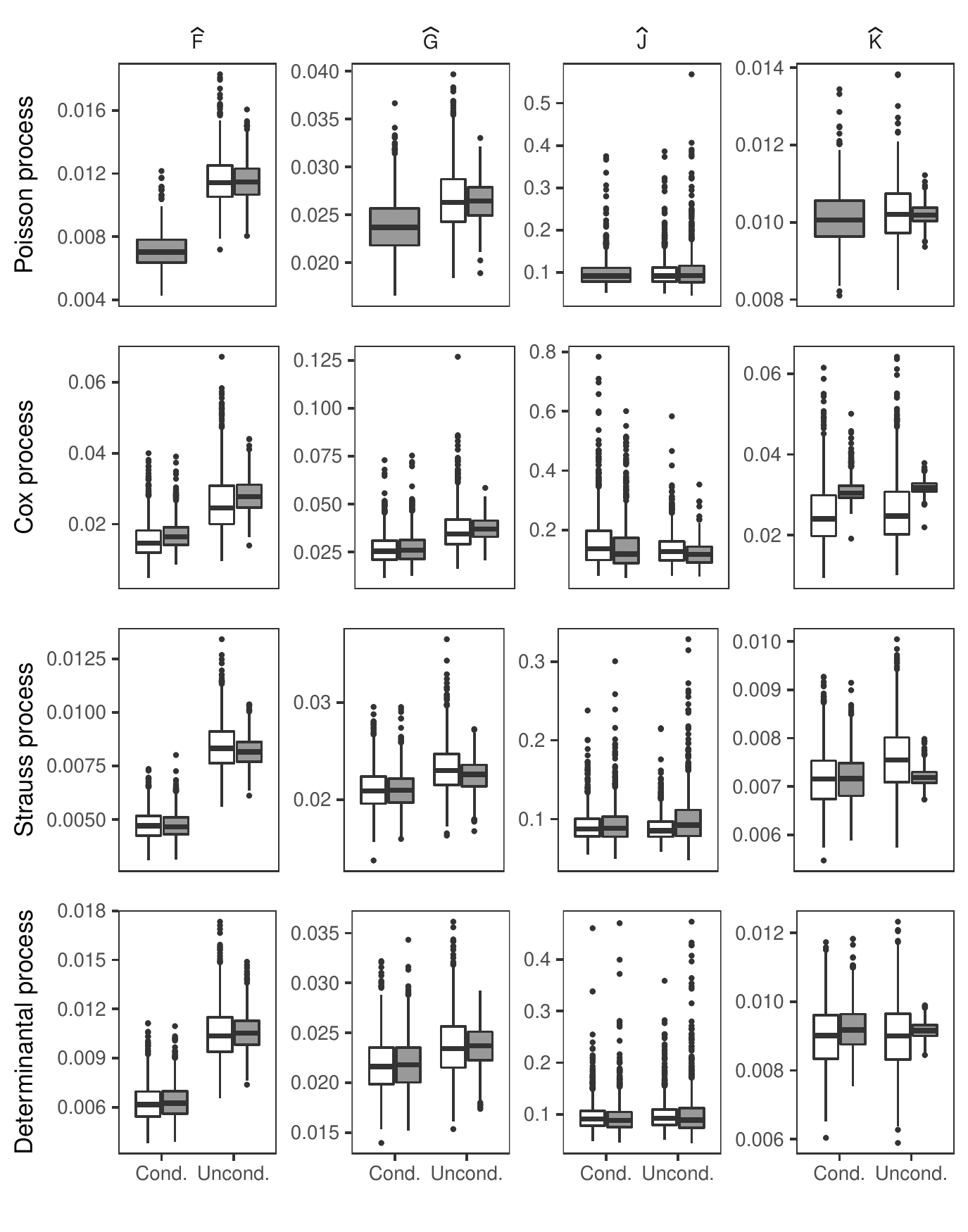}
\caption{Boxplots for the area between 95\% global envelope curves for the functional summary statistic stated at the top of each column. The results were obtained from the simulation study described in the text. The grey level indicates whether the true parameters (dark grey) or fitted parameters (white) were used in the simulations. Whether simulations were made conditional on the number of points is stated at the bottom of each column. The type of model under consideration is stated to the left of each row.}
\label{fig:GET_width}
\end{figure}

Figure~\ref{fig:GET_width} shows boxplots of the approximated area of the envelopes. We see that it generally makes little difference in the area of the envelopes whether the parameters are fitted from data or not, except for $\hat{K}$ where there is less variation in the width of the envelopes when using the true parameters especially in the unconditional case.  We also see that for $\hat K$ and $\hat J$ it makes no real difference in the area of the envelopes to condition on the number of points either, but for $\hat G$ and especially $\hat F$ the envelopes are in general more narrow in the conditional case.

\begin{figure}[htp!]
\centering
\includegraphics[scale=1]{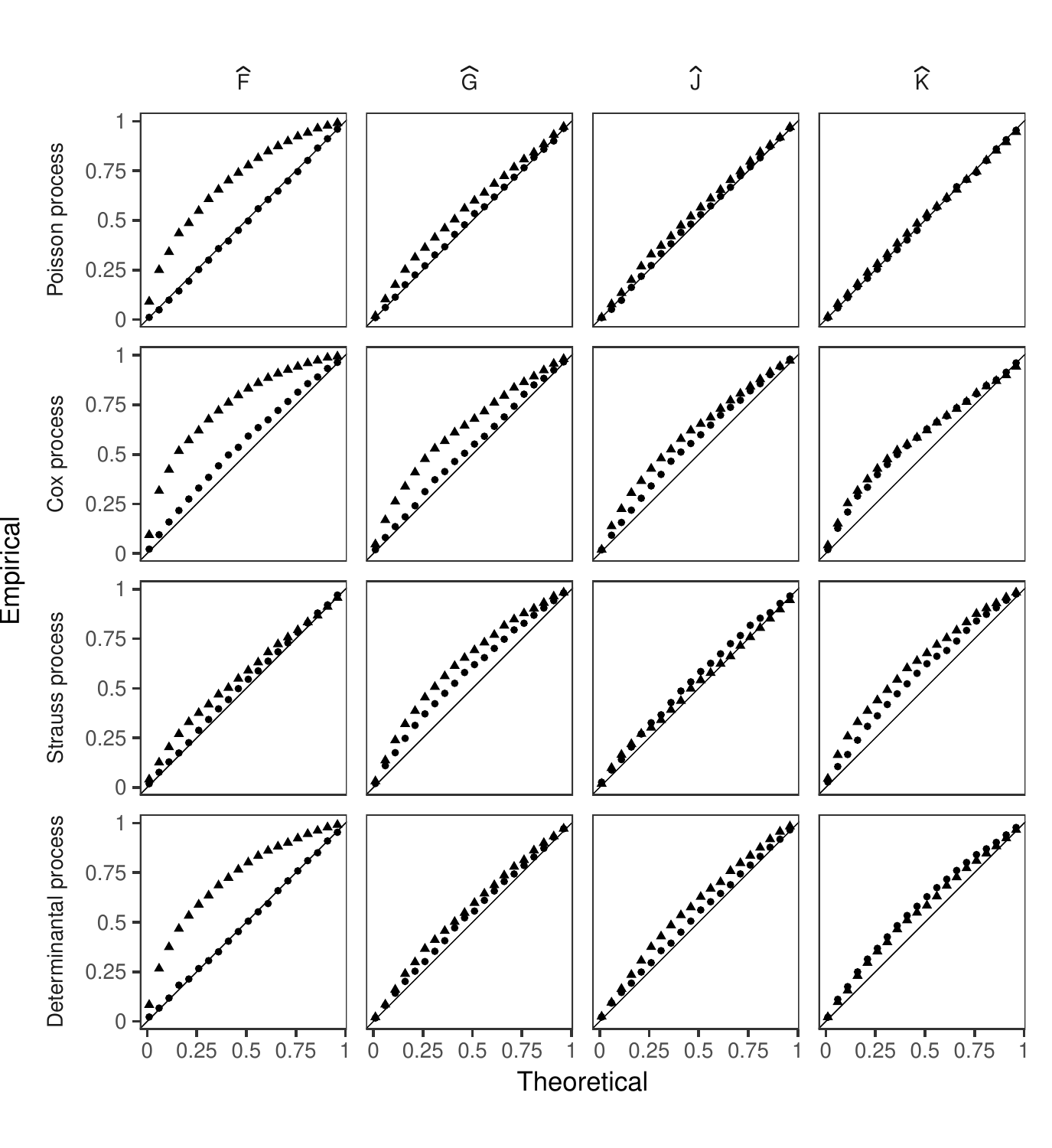}
\caption{Quantile-quantile plots comparing the empirical distributions of the $p$-values of global envelope tests to a uniform distribution on $[0,1]$. The considered $p$-values are the results of the simulation study described in the text. The type of point indicates whether the global envelope tests were based on conditional (dots) or unconditional (triangles) simulations under the fitted model. The functional summary statistic used is stated at the top of each column, the type of model is stated to the left of each row and the straight line is the identity line. 
}
\label{fig:GET_pvalues}
\end{figure}

Figure~\ref{fig:GET_pvalues} shows quantile-quantile plots comparing the distributions of the $p$-values of the global envelope tests for each fitted model to a uniform distribution on $[0,1]$. We see that some of the tests are too conservative, which may be because the null hypothesis is composite except in the conditional case of the Poisson process. Using $\hat{F}$ overall gives very conservative tests in the unconditional case, especially for the Poisson, Cox and determinantal point process, and this behaviour is corrected very well by conditioning on the number of points. For $\hat{G}$, it makes little difference to condition on the number of points except in the case of the Cox process where it corrects the conservativeness in the unconditional case very well. For both $\hat{J}$ and $\hat{K}$, it makes little to no difference whether we condition on the number of points or not, and in all cases the distributions of the $p$-values are in good agreement with the uniform distribution even though we also see some slight conservativeness in some tests. Since $\hat{K}$ was used to fit the parameters of the Cox and determinantal point processes, we usually do not want to use $\hat{K}$ for model validation as well. 

Moreover, to see the effect of increasing the intensity, we made a simulation study for the stationary Poisson process with $\rho = 200$. Our conclusions remained the same and the only real difference was that all envelopes were in general more narrow, which was to be expected since the summary statistics varies less when we observe more points. However, the simulation studies for the remaining three models are time consuming, and it will be even worse if we increase the intensity, but we do not believe the conclusions will be much different as we have indeed established for the example of a Poisson process.

\section{Conditional estimation}
\label{s:est}

In this section, we investigate the possibility to estimate parameters in spatial point process models conditional on the number of points. We consider Cox, Gibbs, and determinantal point processes and in each case discus whether conditional estimation offers any advantages over the unconditional approach. There is of course no reason to consider conditional estimation in the case of a stationary Poisson process since the conditional case is a binomial point process with no unknown parameters.

\subsection{Cox processes}

Parametric models for a stationary Cox process $X$ driven by a random field $Z$ on $\mathbb R^d$ are usually of the form $Z(u)=\rho R(u)$ where $R=\{R(u)\}_{u\in \mathbb R^d}$ is a non-negative unit-mean stochastic process following a parametric model with a parameter $\psi$ so that $(\rho,\psi)$ has range $(0,\infty)\times\Psi$ for some set $\Psi\subseteq\mathbb R^p$. The process $X_W$ then has a density
\begin{equation}\label{e:dc1}
f(x)=\mathrm \rho^{n}\mathrm E\left[R(x_1)\ldots R(x_n)\exp\left\{|W|-\rho\int_W R(u)\,\text{d}u\right\}\right]
\end{equation}
with respect to a Poisson process of intensity 1 and restricted to $W$. In general, this density is not expressible on closed form except for rather trivial cases, and so likelihood based inference is difficult although a missing-data Markov chain Monte Carlo approach can be used for approximate maximum likelihood estimation, see \cite{textbook}. Since second-order moments of the counts $N(B)$ are often expressible on closed form, moment-based and composite likelihood estimation procedures are usually preferred, see \cite{Annual} and the references therein.

When conditioning on $N(W)=n$, any ordering $(X_1,\ldots,X_n)$ of the $n$ points in $X_W$ has a density with respect to Lebesgue measure on $W^n$ which is proportional to the right hand side in \eqref{e:dc1}. Hence, it is also not expressible on closed form and depends on both $\rho$ and $\psi$. Furthermore, moment-based estimation is no longer possible. Consequently, we do not see any advantage in conditioning on the number of points. 

\subsection{Gibbs processes}\label{s:Gibbs_est}

Suppose that $X$ is a stationary Gibbs point process with known interaction radius $R>0$ (details for this general setting are provided in Appendix~\ref{app:Gibbs_def_sim}. Further, suppose that $X_W=x$ has been observed, and let $x_{W_{\ominus R}}= \{x_1,\ldots,x_m\}$. We define the $R$-close neighbourhood to $W_{\ominus R}$ by $\partial W_{\ominus R}\coloneqq (W_{\ominus R})_{\oplus R}\setminus W_{\ominus R}$ and base inference on the conditional distribution of $X_{W_{\ominus R}}$ given $X_{\partial W_{\ominus R}}=x_{\partial W_{\ominus R}}$. Thereby, we account for edge effects due to the unobserved points in $X_{\mathbb{R}^d\setminus W}$ because $X_{W_{\ominus R}}$ and $X_{\mathbb{R}^d\setminus W}$ are independent conditioned on $X_{\partial W_{\ominus R}}=x_{\partial W_{\ominus R}}$. Below, we discuss reasons and consequences of further conditioning on $N(W_{\ominus R})=m$.

For specificity and simplicity, let us think of $X$ as the stationary Strauss process given in Section~\ref{s:models_sim}. Then the likelihood function is of exponential family form with canonical parameter $(\log\beta,\log\gamma)$ and minimal sufficient statistic $(n(x_{W_{\ominus R}}),s(x_{W_{\ominus R}} \cup x_{\partial W_{\ominus R}}))$. However, estimation of $(\beta,\gamma)$ is complicated by the fact that the normalizing constant is not expressible on closed form for $\gamma<1$. Often, the interaction parameter $\gamma$ is of main interest; if we also condition on $N(W_{\ominus R})=m$, we obtain a likelihood function which only depends on $\gamma$. The normalizing constant of that likelihood function is also not expressible on closed form for $\gamma<1$, but it is simpler to approximate, cf.\ Appendix~\ref{app:MLEGibbs}. In particular, as noticed at the end of Appendix~\ref{app:MLEGibbs}, simulations and  computations will be faster when conditioning on $N(W_{\ominus R})=m$. However, it is still slower and more difficult than using maximum pseudo-likelihood estimation, which will be described next. 

Maximum pseudo-likelihood estimation is known to be a fast and often reliable alternative to maximum likelihood estimation, see e.g.\ \cite{PL1}, \cite{PL2}, and the references therein. The definition of the pseudo-likelihood function depends on the context and is specified in Appendix~\ref{app:MPLEGibbs} both with and without conditioning on $N(W_{\ominus R})=m$ and in a way which accounts for edge effects. In the conditional case, we consider Besag's original pseudo-likelihood function \citep{Besag}; in the unconditional case, we consider Besag's extension to spatial point processes \citep{Besag2,PL1}. The computational advantage of using the pseudo-likelihood functions is that they do not depend on the intractable normalizing constant which appears in maximum likelihood estimation. 

We wanted to investigate whether it makes a real difference in the maximum pseudo-likelihood estimate of $\gamma$ to condition on the number of points. In order to do this, we considered 1000 simulations on the unit square of a stationary Strauss process when $\beta = 200$, $R = 0.05$, and $\gamma$ was sampled uniformly in the interval $[0.01, 1]$. For each simulation, we calculated the maximum pseudo-likelihood estimate of $(\beta,\gamma)$ with the function \texttt{exactMPLEstrauss} in \texttt{spatstat}, which computes the maximum pseudo-likelihood estimate to a high accuracy. We also found the maximum pseudo-likelihood estimate in the conditional case by implementing a function where we used the same numerical methods for optimization and evaluation of integrals as in \texttt{exactMPLEstrauss}. In both cases, we let $R=0.05$ be given. The average absolute difference between the estimates of $\gamma$ obtained with the two methods was $0.005$, and the largest absolute difference was $0.01$. So, it makes very little difference to condition on the number of points. Furthermore, the computations for the pseudo-likelihood function in the conditional case may be more cumbersome since more integrals have to be evaluated, cf.\ Appendix~\ref{app:MPLEGibbs}. Therefore, there is no apparent reason to use the more complicated pseudo likelihood method of the conditional case.

\subsection{Determinantal point processes}

Parametric models for stationary determinantal point processes with intensity $\rho>0$ are specified by a parametric class of kernel functions which are usually of the form $C_{\theta}(u,v) = \rho R_\psi(u-v)$ where $\theta=(\rho,\psi)$ and $(u,v)\mapsto R_\psi(u-v)$ is a (complex) correlation function.  
Under weak assumptions, e.g.\ that $C_\theta$ is a continuous complex covariance function, the kernel restricted to $W\times W$ has a spectral representation
\begin{equation*}
C_{\theta}(u,v)=\sum_{i=1}^\infty\lambda_i\phi_i(u)\overline{\phi_i(v)},\qquad u,v\in W,
\end{equation*}
where $\{\phi_i\}_{i=1,2,\ldots}$ is an orthonormal basis for the $L^2$-space of square-integrable complex functions on $W$ and $\lambda_1,\lambda_2,\ldots$ are corresponding eigenvalues. We have $\lambda_i=\rho\lambda_i'$ where $(\phi_i,\lambda_i')$ depends only on $\psi$ and the condition $0\le\lambda_i\le1$ is needed to ensure existence of the process. 
Therefore, the parameter space $\Psi_\rho$ of $\psi$ will depend on the value of $\rho$; briefly speaking, there is a trade-off between intensity and repulsion, and  the set $\Psi_\rho$ decreases as $\rho$ increases, cf.\ \cite{DPP}. For parameter estimation based on maximum likelihood and moment-based methods, see \cite{DPP}. 

Conditioned on $N(W)=n$, any ordering $(X_1,\ldots,X_n)$ of the $n$ points in $X_W$ has probability density function 
\begin{equation*}
\frac{\sum_{i_1<\cdots<i_n}\left(\prod_{j=1}^n\lambda_{i_j}\right)\left(\prod_{j\not\in\{i_1,\ldots,i_n\}}(1-\lambda_j)\right)\frac{1}{n!}\mathrm{det}\bigg\{\sum_{k=1}^n\phi_{i_k}(x_i)\overline{\phi_{i_k}(x_j)}\bigg\}_{i,j=1,\ldots,n}}{\sum_{i_1<\cdots<i_n}\left(\prod_{j=1}^n\lambda_{i_j}\right)\left(\prod_{j\not\in\{i_1,\ldots,i_n\}}(1-\lambda_j)\right)}
\end{equation*}
for $(x_1,\ldots,x_n)\in W^n$. For parametric models as considered above, it follows that the conditional distribution of $X_W$ given $N(W)=n$ depends on both $\rho$ and $\psi$ in a complicated way; in fact it is more complicated than the likelihood in the unconditional case, cf.\ \cite{DPP}. Therefore, we do not see any advantage in conditioning on $N(W)=n$ when making parameter estimation. 

\section{Concluding remarks}\label{s:conclusion}

It is worth repeating that any claimed model for the number of points cannot be justified based on just one realization. However, in case of Cox, Gibbs and determinantal point processes, it is convenient for computational reasons to assume a distribution for the number of points when estimating parameters. 
 
Regarding global envelope tests based on $\hat K$ or $\hat J$, it made little or no difference in our simulation study whether we condition on the number of points. However, when we instead used $\hat F$ or $\hat G$, conditioning on the number of points gave more narrow envelopes and hence stronger tests, and it corrected for conservativeness in the tests. 

Global envelopes and tests are usually calculated from simulations of a single point process model. For a composite null hypothesis, it is possible to make an adjusted global envelope test but at the expense of many simulations\citep{GETinR}. Conditioning on the number of points may offer an alternative which requires fewer simulations, but whether this will be faster in practice depends on the actual speed of the simulation procedures. We leave this for future research.

Regarding conditional estimation, we concluded that it is impractical and offers no clear advantage to consider conditional estimation for Cox and determinantal point process models. For Gibbs point process models, it simplifies maximum likelihood estimation but complicates the more commonly used and faster method of maximum pseudo-likelihood estimation where there is also little difference in the estimates achieved with and without conditioning. We therefore overall have found no apparent reason to use conditional estimation. 

\section*{Acknowledgements} The research of the authors was supported by The Danish Council for Independent Research | Natural Sciences, grant DFF -- 7014-00074 `Statistics for point processes in space and beyond'.

\bibliographystyle{plainnat} 
\bibliography{bib}

\begin{thebibliography}{}

\bibitem[Baddeley et~al., 2014]{PL2}
Baddeley, A., Coeurjolly, J.-F., Rubak, E., \& Waagepetersen, R. (2014).
\newblock Logistic regression for spatial {G}ibbs point processes.
\newblock {\em Biometrika}, 101, 377--392.

\bibitem[Baddeley et~al., 2015]{spatstat}
Baddeley, A., Rubak, E., \& Turner, R. (2015).
\newblock {\em Spatial Point Patterns: Methodology and Applications with {R}}.
\newblock Boca Raton: Chapman and Hall/CRC Press.

\bibitem[Barndorff-Nielsen, 1978]{Barndorff}
Barndorff-Nielsen, O.~E. (1978).
\newblock {\em Information and Exponential Families in Statistical Theory}.
\newblock New York: Wiley.

\bibitem[Besag, 1975]{Besag}
Besag, J. (1975).
\newblock Statistical analysis of non-lattice data.
\newblock {\em The Statistician}, 24, 179--195.

\bibitem[Besag, 1977]{Besag2}
Besag, J. (1977).
\newblock Some methods of statistical analysis for spatial data.
\newblock {\em Bulletin of the International Statistical Institute}, 47,
  77--91.

\bibitem[Georgii, 1988]{Georgii:88}
Georgii, H.-O. (1988).
\newblock {\em Gibbs Measures and Phase Transition}.
\newblock Berlin: Walter de Gruyter.

\bibitem[Geyer \& Møller, 1994]{Geyer-Moeller}
Geyer, C.~J. \& Møller, J. (1994).
\newblock Simulation procedures and likelihood inference for spatial point
  processes.
\newblock {\em Scandinavian Journal of Statistics}, 21, 359--373.

\bibitem[Jensen \& Møller, 1991]{PL1}
Jensen, J.~L. \& Møller, J. (1991).
\newblock Pseudolikelihood for exponential family models of spatial point
  processes.
\newblock {\em Annals of Applied Probability}, 1, 445--461.

\bibitem[Lavancier et~al., 2015]{DPP}
Lavancier, F., Møller, J., \& Rubak, E. (2015).
\newblock Determinantal point process models and statistical inference.
\newblock {\em Journal of Royal Statistical Society: Series B (Statistical
  Methodology)}, 77, 853--877.

\bibitem[Macchi, 1975]{Macchi}
Macchi, O. (1975).
\newblock The coincidence approach to stochastic point processes.
\newblock {\em Advances in Applied Probability}, 7, 83–122.

\bibitem[Møller et~al., 1998]{LGCP}
Møller, J., Syversveen, A.~R., \& Waagepetersen, R.~P. (1998).
\newblock Log {G}aussian {C}ox processes.
\newblock {\em Scandinavian Journal of Statistics}, 25, 451--482.

\bibitem[Møller \& Waagepetersen, 2017]{Annual}
Møller, J. \& Waagepetersen, R. (2017).
\newblock Some recent developments in statistics for spatial point patterns.
\newblock {\em Annual Review of Statistics and Its Applications}, 4, 317--342.

\bibitem[Møller \& Waagepetersen, 2004]{textbook}
Møller, J. \& Waagepetersen, R.~P. (2004).
\newblock {\em Statistical Inference and Simulation for Spatial Point
  Processes}.
\newblock Boca Raton: Chapman and Hall/CRC.

\bibitem[Moran, 1952]{Moran}
Moran, P. (1952).
\newblock A characteristic property of the {P}oisson distribution.
\newblock {\em Mathematical Proceedings of the Cambridge Philosophical
  Society}, 48, 206--207.

\bibitem[Mrkvi{\v{c}}ka et~al., 2018]{GET2018}
Mrkvi{\v{c}}ka, T., Myllym{\"a}ki, M., J{\'i}lek, M., \& Hahn, U. (2018).
\newblock A one-way {ANOVA} test for functional data with graphical
  interpretation.
\newblock Available at arXiv:1612.03608.

\bibitem[Myllym{\"a}ki \& Mrkvi{\v{c}}ka, 2019]{GETinR}
Myllym{\"a}ki, M. \& Mrkvi{\v{c}}ka, T. (2019).
\newblock {GET}: Global envelopes in {R}.
\newblock Available at arXiv:1911.06583.

\bibitem[Myllym{\"a}ki et~al., 2017]{GET2017}
Myllym{\"a}ki, M., Mrkvi{\v{c}}ka, T., Grabarnik, P., Seijo, H., \& Hahn, U.
  (2017).
\newblock Global envelope tests for spatial processes.
\newblock {\em Journal of the Royal Statistical Society: Series B (Statistical
  Methodology)}, 79, 381--404.

\bibitem[{R Core Team}, 2019]{R}
{R Core Team} (2019).
\newblock {\em R: A Language and Environment for Statistical Computing}.
\newblock R Foundation for Statistical Computing, Vienna, Austria.

\bibitem[Ripley, 1977]{Ripley:77}
Ripley, B.~D. (1977).
\newblock Modelling spatial patterns.
\newblock {\em Journal of Royal Statistical Society: Series B (Statistical
  Methodology)}, 39, 172--192.

\bibitem[Ripley, 1979]{ripley79}
Ripley, B.~D. (1979).
\newblock Algorithm {AS} 137: simulating spatial patterns: dependent samples
  from a multivariate density.
\newblock {\em Journal of the Royal Statistical Society. Series C (Applied
  Statistics)}, 28, 109--112.

\bibitem[Ripley, 1988]{Ripley:88}
Ripley, B.~D. (1988).
\newblock {\em Statistical Inference for Spatial Processes}.
\newblock Cambridge: Cambridge University Press.

\bibitem[Wickham, 2016]{ggplot}
Wickham, H. (2016).
\newblock {\em ggplot2: Elegant Graphics for Data Analysis}.
\newblock Springer-Verlag New York.

\end{thebibliography}

\appendix
\section{Details regarding the simulation study in Section~\ref{s:simstudy_GET}}\label{app:simdetails}
The following gives some practical and technical details regarding the simulation study described in Section~\ref{s:setup}.

 When calculating functional summary statistics, we always used the default range of $r$-values in \texttt{spatstat}. For details on simulation procedures, especially conditional simulation, see Appendix~\ref{app:models}. When fitting parameters, we always did it without conditioning on the number of points since we argue in Section~\ref{s:est} that there is no apparent reasons to use conditional estimates. We used the natural estimate $n(x)$ of the intensity in the case of the Poisson process, used the method of minimum contrast estimation based on Ripley's $K$-function (see \cite{spatstat}) in the cases of the log-Gaussian Cox process and the Gaussian determinantal point process and used the method of profile maximum pseudo likelihood in the case of the Strauss process (where we considered 41 equally spaced values of $R$ in the interval $(0.03, 0.07)$). When fitting the parameters of the log-Gaussian Cox process, we found that sometimes the scale parameter $\delta$ was seriously overestimated, which caused the conditional simulation procedure to be extremely slow. Therefore, we decided only to use realizations of the log-Gaussian Cox process M2 where the fitted scale parameter was below $0.3$, which left $953$ realizations in the simulation study. When fitting parameters in the case of the Gaussian determinantal point process, we found that $\kappa$ was seriously underestimated for a few realizations, which either slowed down the simulation procedures considerably or caused them to fail. We therefore excluded realizations with a fitted value of $\kappa$ less than 0.001 after which 996 realizations remained.

\section{Stationary point process models and conditional si\-mu\-la\-tion}\label{app:models}

In the following, we describe some popular classes of spatial point process models: Poisson, Cox, Gibbs and determinantal point processes. It is well known how to simulate such point processes within the compact observation window $W$ without conditioning on the number of points; for Poisson, Cox and Gibbs point processes, see \citet{textbook} and the references therein, and for determinantal point processes, see \citet{DPP}. It is also well-known how to simulate Poisson processes when we condition on the number of points falling in $W$. For finite Gibbs point processes defined on $W$ (and thus not the infinite stationary Gibbs processes which we consider), \citet{ripley79} described how to make simulations conditioned on the number of points but without accounting for edge effects. Below, we suggest methods for simulation of Cox, infinite Gibbs and determinantal point processes conditioned on $N(W)$ and where we account for edge effects when needed (the case of infinite Gibbs point processes).  

\subsection{Poisson processes}

Suppose that $\rho:\mathbb{R}^d\to [0,\infty)$ satisfies that $\nu(B)\coloneqq\int_{B}\rho(u)\,\mathrm du<\infty$ for bounded (Borel) sets $B\subset \mathbb{R}^d$. A point process $X$ is a Poisson process with intensity $\rho$ if for any $B\subset \mathbb{R}^d$ with $0<\nu(B)<\infty$, $N(B)$ follows a Poisson distribution with mean $\nu(B)$, and for any $n\in\mathbb{N}$, conditioned on $N(B)=n$, the $n$ points in $X_B$ are independent and identically distributed with a density proportional to $\rho$. If $\rho$ is constant, $X$ is a stationary Poisson process. It is usually easy to simulate a Poisson process on $W$ conditioned on $n$, in particular in the stationary case where the $n$ points are just independent and uniformly distributed on $W$.  

\subsection{Cox processes}
\label{app:Cox_def_sim}

Let $Z=\{Z(u)\}_{u\in \mathbb R^d}$ be a non-negative stochastic process such that (almost surely) for every bounded set $B\subset\mathbb R^d$, $\int_B Z(u)\,\mathrm du$ exists and is finite. Assuming $X$ conditioned on $Z$ is a Poisson process with intensity function $Z$, we call $X$ a Cox process driven by $Z$. When $Z$ is stationary with finite mean, $X$ is stationary with intensity $\rho=\mathrm EZ(u)$ for any $u\in\mathbb R^d$.  

For simulating $X_W$ conditioned on $N(W)=n$, we can use the following simple accep\-tance-rejection sampling procedure. Defining $Z_W\coloneqq \{Z(u)\}_{u\in \mathbb W}$, repeat simulating a realization $Z_W=z$ and an independent uniform variable $U$ on $[0,1]$ until $$U\le\frac{1}{n!}\left(\int_W z(u)\,\mathrm du\right)^n\exp\left(-\int_W z(u)\,\mathrm du\right).$$ Then, simulate $n$ points from $W$ independently with a density proportional to $z$. For a log-Gaussian Cox process, $Z = \exp(Y)$ with $Y$ a Gaussian random field defined on $\mathbb R^d$. In this case, the conditional simulation procedure can be refined by considering a subdivision of $W$ as described in \citet{LGCP}.

\subsection{Gibbs point processes}
\label{app:Gibbs_def_sim}

The definition of a stationary Gibbs point process is rather technical. In the following definition using a local specification, we omit for simplicity not only measure theoretical details but also technical conditions ensuring existence of the process; for such details, we refer instead to \cite{Georgii:88} or the review in \cite{textbook}. 

We need the following notation. Let $R$ be a given non-negative number and $B\subset\mathbb R^d$ a bounded set. As a reference measure we consider the stationary Poisson process on $\mathbb R^d$ with intensity 1 and denote its distribution $\mu$. The restriction of $\mu$ to $B$ is denoted $\mu_B$. The $R$-close neighbourhood to $B$ is defined by $\partial B\coloneqq B_{\oplus R}\setminus B$. Let $\Omega_{{\mathrm{fin}}}\subset\Omega$ be the set of all finite subsets of $\mathbb R^d$. Consider functions $s:\Omega_{{\mathrm{fin}}}\mapsto\mathbb R^p$ and $a:\Omega_{{\mathrm{fin}}}\mapsto[0,\infty)$ satisfying the following conditions:  $a$ is hereditary, that is, $a(x)>0$ implies $a(y)>0$ for $y\subset x$; $a$ and $s$ are invariant under translations in $\mathbb R^d$ (this will be needed to ensure that $X$ is stationary); and for every $x\in\Omega_{{\mathrm{fin}}}$ and every $u\in\mathbb R^d\setminus x$, setting 
\[a(x,u)\coloneqq a(x\cup\{u\})/a(x),\qquad s(x,u)\coloneqq s(x\cup\{u\})-s(x),\]  
then 
\begin{equation}\label{e:M2}
a(x,u)=a(x\cap b(u,R),u),\qquad s(x,u)=s(x\cap b(u,R),u).
\end{equation}
Finally, we consider a parameter $\theta=(\beta,\psi)$ with $\beta>0$ and $\psi\in\Psi$ so that $\Psi\subseteq\mathbb R^p$ and $\Theta\coloneqq (0,\infty)\times\Psi$ is the parameter space. 

Now, we consider a local specification which is given by the following assumptions (i)--(ii). For every bounded set $B\subset\mathbb R^d$ and every $\theta\in\Theta$, we have: 
(i) $X_B$ and $X_{\mathbb{R}^d\setminus B_{\oplus R}}$ are conditionally independent given $X_{\partial B}$; (ii) for ($\mu_{\partial B}$ almost) every finite $x_{\partial B}\subseteq \partial B$,  
$X_B$ conditioned on $X_{\partial B}=x_{\partial B}$ has a density with respect to $\mu_B$ given by 
\begin{equation}
f_{B,\theta}(x_B\,|\,x_{\partial B})=\label{e:cond1} a(x)\beta^{n(x_B)}\exp(\psi^T s(x))/c_{B,\theta}(x_{\partial B})\qquad \mbox{for finite }x_B\subseteq B,
\end{equation}
where $x=x_B\cup x_{\partial B}$, $s(x)$ and $\psi$ are viewed as column vectors, $\psi^T$ is the transpose of $\psi$ and $c_{B,\theta}(x_{\partial B})$ is a normalizing constant. Then, we call $X$ an infinite Gibbs (or Markov) point process with parameter $\theta$ and interaction radius $R$. 

Above, we have imposed more structure than usual when defining a Gibbs point process: the assumption in \eqref{e:M2} is a local Markov property, which is in line with the spatial Markov properties specified in (i), and in \eqref{e:cond1}, we have imposed an exponential family structure. In applications, we typically interpret $\psi$ as an interaction parameter, and $\beta$ usually controls the intensity, which also depends on $\psi$. For example, a stationary Strauss process has $a=1$, $\Psi=(-\infty,0]$ and
\[s(x)=s_R(x)\coloneqq \sum_{i<j}\mathbbm 1(\|x_i-x_j\|\le R)\qquad\mbox{for }x=\{x_1,\ldots,x_n\}\in\Omega_{{\mathrm{fin}}}.
\] 
%which is the number of $R$-close pairs in $x$.

For conditional simulation of $X_W$ given $N(W)=n$, we consider an extended compact region $W_{\mathrm{ext}}\supseteq W_{\oplus R}$ in order to account for edge effects. Then, we let $Y$ denote a process on $W_{\mathrm{ext}}$ with a density with respect to $\mu_{W_{\mathrm{ext}}}$ given by 
\begin{equation*}
f_{\theta}(x)=a(x)\beta^{n(x)}\exp(\psi^{T}s(x))/c_{\theta}, \quad x\subset W_{\mathrm{ext}},
\end{equation*}
assuming the normalising constant $c_{\theta}$ is finite. Simulations of $Y_W$ will then be approximate simulations of $X_W$ if $W_{\mathrm{ext}}$ is large enough. Let $(Y_1,\ldots,Y_n)$ be an arbitrary ordering of the $n$ points in $Y_W$, and define $A\coloneqq W_{\mathrm{ext}}\setminus W$. The conditional density of $(Y_1,\ldots,Y_n)$ given both $Y_A=y_A$ and $n(Y_W)=n$ is
\begin{equation}\label{e:sim1}
f_\psi(y_1,\ldots,y_n\,|\,y_{\partial W},n)\propto a(\{y_1,\ldots,y_n\}\cup y_{\partial W}) \exp(\psi^Ts(\{y_1,\ldots,y_n\}\cup y_{\partial W}))
\end{equation}
with respect to Lebesgue measure on $W^n$, and it only depends on $y_A$ through $y_{\partial W}$ and on $\theta$ through $\psi$. Moreover, the conditional density of $Y_A$ given both $(Y_1,\ldots,Y_n)=(y_1,\ldots,y_n)$ and $n(Y_W)=n$ depends only on $(y_1,\ldots,y_n)$ through those $y_i\in\partial A$. It is
\begin{equation}\label{e:sim2}
f_\theta(y_A\,|\,\{y_i:y_i\in\partial A\},n)\propto a(y_A\cup \{y_i:y_i\in\partial A\}) \beta^{n(y_A)}\exp(\psi^Ts(y_A\cup \{y_i:y_i\in\partial A\}))
\end{equation}
with respect to $\mu_A$. 

We use a Gibbs within Metropolis-Hastings algorithm where we alternate between simulating from \eqref{e:sim1} and \eqref{e:sim2}. In case of \eqref{e:sim1}, a single point updating Metropolis algorithm  is used (specifically, Algorithm~7.2 in  \cite{textbook} where a proposal consists in replacing a uniformly selected existing point $y_i$ with another point $y_i'$ which is uniformly distributed on $W$). In case of \eqref{e:sim2}, we use the birth-death Metropolis-Hastings algorithm in \cite{Geyer-Moeller} (Algorithm~7.4 in \cite{textbook}).

In practice, it is necessary to choose an appropriate burnin when using Markov chains for simulation. For our simulations, based on various trace plots, we chose to use a burnin of 4000 and 1000 iterations for unconditional and conditional simulation of Strauss processes, respectively. It is also necessary to decide what $W_{\mathrm{ext}}$ should be. We used the default in the function \texttt{rmh} from \texttt{spatstat}, which was also the function we used to make unconditional simulations.

\subsection{Determinantal point processes}
\label{app:DPP_def_sim}

We say that $X$ is a determinantal point process with kernel $C:\mathbb R^d\times\mathbb R^d\mapsto C$ if for $n=1,2,\ldots$ and any pairwise disjoint bounded (Borel) sets $A_1,\ldots,A_n\subset\mathbb R^d$,  
\begin{equation}\label{e:DPPmoments}
\mathrm E[N(A_1)\cdots N(A_n)]=\int_{A_1}\cdots\int_{A_n}\mathrm{det}\{C(u_i,u_j)\}_{i,j=1,\ldots,n}\,\mathrm du_1\cdots\,\mathrm du_n
\end{equation}
where $\mathrm{det}\{C(u_i,u_j)\}_{i,j=1,\ldots,n}$ is the determinant of the $n\times n$ matrix with $(i,j)$'th entry $C(u_i,u_j)$. For $1<k<n$ we have
\[{\mathrm{cov}}(N(A_1)\cdots N(A_k),N(A_{k+1})\cdots N(A_n))\le0,\]
and the process is therefore said to be repulsive at all scales. It is often assumed that $C$ is a continuous complex covariance function so that $C(u,v)=C_0(u-v)$ is translation invariant; we also make that assumption, which implies that $X$ is stationary. For details on the many nice properties of determinantal point processes including those given below, see \cite{DPP} and the references therein.

The kernel restricted to $W\times W$ has a spectral representation
\begin{equation}\label{e:spectral}
C(u,v)=\sum_{i=1}^\infty\lambda_i\phi_i(u)\overline{\phi_i(v)},\qquad u,v\in W
\end{equation}
where $\{\phi_i\}_{i=1,2,\ldots}$ is an orthonormal basis for the $L^2$-space of square-integrable complex functions on $W$ and each $\lambda_i$ is an eigenvalue. Existence of the determinantal point process restricted to $W$ is equivalent to that each $\lambda_i\in[0,1]$ and $\sum_{i=1}^\infty\lambda_i<\infty$. Now, let $B_1,B_2,\ldots$ be independent Bernoulli variables with parameters $\lambda_1,\lambda_2,\ldots$. Then, $X_W$ is distributed as a determinantal point process on $W$ with kernel
\begin{equation*}
K(u,v)\coloneqq\sum_{i=1}^\infty B_i\phi_i(u)\overline{\phi_i(v)},\qquad u,v\in W,
\end{equation*}
and $N(W)$ is distributed as $\sum_{i=1}^\infty B_i$, which is finite with probability 1. It can be shown that $X_{W}$ conditional on a realization $b_1,b_2,\ldots$ of $B_1,B_2,\ldots$ has $n=\sum_{i=1}^\infty b_i$ points, and any ordering $(X_1,\ldots,X_n)$ of these points has probability density function
\[p_{i_1,\ldots,i_n}(x_1,\ldots,x_n)\coloneqq \frac{1}{n!}\mathrm{det}\bigg\{\sum_{k=1}^n\phi_{i_k}(x_i)\overline{\phi_{i_k}(x_j)}\bigg\}_{i,j=1,\ldots,n}\]
on $W^n$ where $i_1,\ldots,i_n$ are the indices for which the Bernoulli variables are 1. It is well-known how to make simulations from this distribution.  

Apart from special cases, only the existence of the spectral representation but not its exact form is known in practice. \cite{DPP} provided efficient ways of obtaining a useful approximation. 

For conditional simulation of $X_W$ given $N(W)=n$, we first simulate $B_1,B_2,\ldots$ conditional on $N(W)=n$ and hereafter simulate $X_W$ conditional on $B_1,B_2,\ldots$, which then does not depend on $N(W)$. Since it is well known how to simulate from the density $p_{i_1,\ldots,i_n}$, it suffices to discuss how to simulate $B_1,B_2,\ldots$ conditioned on $N(W)=n$. Define $I_0\coloneqq 0$ and $I_k\coloneqq \inf\{j>I_{k-1}\,|\,B_j=1\}$ ($k=1,2,\ldots$), setting $\inf\emptyset\coloneqq \infty$. So, $I_k$ is the $k$th time that a Bernoulli variable is 1, that is, $B_i=1$ if $i\in\{I_1,\ldots,I_n\}$ and $B_i=0$ otherwise. We simulate a realization of $(I_1,\ldots,I_n)$ as follows. For any integers $0\le\ell<k$ such that $\prod_{\ell<j<\infty}(1-\lambda_j)<1$, define
 \[p(k\,|\,\ell)\coloneqq \frac{\lambda_k\prod_{\ell<j<k}(1-\lambda_j)}{1-\prod_{\ell<j<\infty}(1-\lambda_j)},\]
where $\prod_{\ell<j<k}(1-\lambda_j)\coloneqq 1$ if $k=\ell+1$. Set $i_0\coloneqq 0$. Then, for $k=1,\ldots,n$ and any integers $i_n>\ldots>i_1\ge1$, 
\[\mathrm P(I_k=i_k\,|\,I_0=i_0,\ldots,I_{k-1}=i_{k-1},I_k<\infty)=\mathrm P(I_k=i_k\,|\,I_{k-1}=i_{k-1},I_k<\infty)=p(i_k\,|\,i_{k-1})\]
and 
\[\mathrm P(I_{n+1}=\infty\,|\,I_0=i_0,\ldots,I_n=i_n)=\prod_{k>i_n}(1-\lambda_k).\]
Hence, a simulation of $(I_1,\ldots,I_n)$ can be generated by the following acceptance-rejection algorithm: (i) For $k=1,\ldots,n$, generate a proposal $i_k$ from the probability mass function $p(\cdot\,|\,i_{k-1})$. (ii) Return $(I_1,\ldots,I_n)=(i_1,\ldots,i_n)$ with probability $\prod_{k>i_n}(1-\lambda_k)$, else go to (i).

For the generation of the proposal in (i), we use inversion sampling: set $\lambda_0\coloneqq 1$ and $F(m|\ell)\coloneqq \sum_{\ell<k\le m}p(k|\ell)$ for integers $0<\ell<k\le m$, which may be computed using the recursion
\[p(\ell+1\,|\,\ell)=\frac{\lambda_{\ell+1}}{1-\prod_{\ell<j<\infty}(1-\lambda_j)},\qquad p(k\,|\,\ell)=p(k-1\,|\,\ell)\frac{\lambda_k(1-\lambda_{k-1})}{\lambda_{k-1}}\qquad\mbox{if }k>\ell+1.\]
Then, generate $U$ from a uniform distribution on $[0,1]$ and return $\inf\{m>\ell\,|\, F(m|\ell)\ge U\}$ as a simulation of the proposal. 

As we need to truncate the infinite products $\prod_{k>i_n}(1-\lambda_k)$ and $\prod_{\ell<j<\infty}(1-\lambda_j)$ by only considering a finite number of eigenvalues, we only get an approximate simulation. For the choice of truncation, we used the default in the function \texttt{simulate.detpointprocfamily} from \texttt{spatstat}, which was also the function we used to make unconditional simulations.

\section{Details for maximum likelihood estimation for\\ Gibbs point processes}\label{app:MLEGibbs}

Estimation of $\theta$ is complicated by the fact that the normalizing constant is in general not expressible on closed form: 
the conditional density of $X_{W_{\ominus R}}$ given $X_{\partial W_{\ominus R}}=x_{\partial W_{\ominus R}}$ has normalizing constant
\begin{equation*}
\begin{aligned}
&c_{W_{\ominus R},\theta}(x_{\partial W_{\ominus R}})=\\
&\quad\sum_{n=0}^\infty \frac{\exp(-|W_{\ominus R}|)}{n!}\int_{W_{\ominus R}}\cdots\int_{W_{\ominus R}} a(y\cup x_{\partial W_{\ominus R}})\beta^{n}\exp(\psi^T  s(y\cup x_{\partial W_{\ominus R}}))\,\mathrm dy_1\cdots\,\mathrm dy_n
\end{aligned}
\end{equation*} 
where $y=\{y_1,\ldots,y_n\}$ and the term for $n=0$ is interpreted as $\exp(-|B|)$. Often, the interaction parameter $\psi$ is of main interest in which case, following \cite{Ripley:77}, it may be reasonable to further condition on $N(W_{\ominus R})=m$. Conditional on $N(W_{\ominus R})=m$, let the random vector $(X_1,\ldots,X_m)$ be any ordering of the $m$ points in $X_{W_{\ominus R}}$, which conditioned on both $X_{\partial W_{\ominus R}}=x_{\partial W_{\ominus R}}$ and $N(W_{\ominus R})=m$ has probability density function
\begin{equation*}
f_{\psi}(x_1,\ldots,x_m\mid x_{\partial W_{\ominus R}},\,m)\coloneqq \frac{a(\{x_1,\ldots,x_m\}\cup x_{\partial W_{\ominus R}})\exp(\psi^Ts(\{x_1,\ldots,x_m\}\cup x_{\partial W_{\ominus R}}))}{c_\psi(x_{\partial W_{\ominus R}},m)}
\end{equation*}
 on $W_{\ominus R}^m$ where 
\begin{align*}
c_\psi(x_{\partial W_{\ominus R}},m)=
\int_{W_{\ominus R}}\cdots\int_{W_{\ominus R}}a(y\cup x_{\partial W_{\ominus R}})\exp(\psi^Ts(y\cup x_{\partial W_{\ominus R}}))\,\mathrm dy_1\cdots \,\mathrm dy_m
\end{align*} 
with $y=\{y_1,\ldots,y_n\}$. This conditional density does not depend on $\beta$ and has $s(x)$ as a sufficient statistic for $\psi$. The $m$-fold integral above may be hard to compute, but at least $c_\psi(x_{\partial W_{\ominus R}},m)$ is simpler than $c_{W_{\ominus R},\theta}(x_{\partial W_{\ominus R}})$. Thus, it seems appealing to condition on both $X_{\partial W_{\ominus R}}=x_{\partial W_{\ominus R}}$ and $N(W_{\ominus R})=m$ when using maximum likelihood estimation. However, in general, $c_{W_{\ominus R},\theta}(x_{\partial W_{\ominus R}})$ considered as a function of $\theta$ cannot be written as a product of two functions with one depending on $\beta$ only and the other depending on $\psi$ only, and so we cannot appeal to one of the known conditioning principles: in general, using a terminology as in \cite{Barndorff}, $n$ will not be an S-ancillary statistic for $\psi$, and $s(x)$ will not be an S-sufficient statistic for $\psi$. 

No matter if we condition on $N(W_{\ominus R})=m$ or not, the likelihood function is log-concave. An approximate maximum likelihood estimate of $\psi$ (and $\beta$ if we do not condition on $N(W_{\ominus R})=m$) can be found by combining simulations with importance sampling to obtain an approximate likelihood function which is log-concave too, see \cite{textbook} and the references therein. Typically, unless $X$ is `close' to a Poisson process, long runs of Markov chains are needed for the simulations. The simulations and the computations will be faster when conditioning on $N(W_{\ominus R})=m$ because the normalizing constant is simpler to approximate and since a single point updating Metropolis algorithm can be used for simulations; in the unconditional case, the more advanced birth-death Metropolis-Hastings algorithm \citep{Geyer-Moeller} is used.

\section{Details for maximum pseudo-likelihood estimation for Gibbs point processes}\label{app:MPLEGibbs}

When defining pseudo-likelihood functions below, we need the Papangelou conditional intensity for the density in \eqref{e:cond1}, which is
\begin{equation*}\label{e:lambda}
\lambda_{B,\theta}(x_B,u\,|x_{\partial B})\coloneqq \frac{f_{B,\theta}(x_B\cup\{u\}\,|\,x_{\partial B})}{f_{B,\theta}(x_B\,|\,x_{\partial B})}=\beta a(x_B\cup x_{\partial B}, u)\exp(\psi^{T}s(x_B\cup x_{\partial B}, u))
\end{equation*}
for $u\in B\setminus x_B$. Furthermore, in order to account for edge effects, we let $B=W_{\ominus R}$ and consider 
\begin{equation}\label{e:papa}
\lambda_{W_{\ominus R},\theta}(x_{W_{\ominus R}},u\,|x_{\partial W_{\ominus R}}) = \lambda_\theta(x,u) \coloneqq \beta a(x, u)\exp(\psi^{T}s(x, u))
\end{equation}
for the observed point pattern $x\subset W$.

First, consider the case where we do not condition on $N(W_{\ominus R})=m$. Then, the log pseudo-likelihood function is
\begin{equation}\label{e:pseudo}
pl(\theta)\coloneqq -\int_{W_{\ominus R}}\lambda_\theta(x,u)\,\mathrm du+\sum_{i=1}^m\log\lambda_\theta(x\setminus\{x_i\},x_i).
\end{equation}
If we fix $\psi$ and insert \eqref{e:papa} into \eqref{e:pseudo}, we see that
\[\hat\beta(\psi)\coloneqq m/\int_{W_{\ominus R}}a(x,u)\exp(\psi^Ts(x,u))\,\mathrm du\]
is the maximum pseudo-likelihood estimate of $\beta$; hence, the profile log pseudo-likelihood function for $\psi$ becomes $pl(\hat\beta(\psi),\psi)$. Assuming that we can interchange differentiation and integration, the pseudo-score function is
\begin{equation}\label{e:score}
s(\psi)\coloneqq\frac{\partial}{\partial\psi}pl(\hat\beta(\psi),\psi)=-m\frac{\int_{W_{\ominus R}}a(x,u)s(x,u)\exp(\psi^Ts(x,u))\,\mathrm du}{\int_{W_{\ominus R}}a(x,u)\exp(\psi^Ts(x,u))\,\mathrm du}
+\sum_{i=1}^m s(x\setminus\{x_i\},x_i),
\end{equation}
which has a negative definite derivative. Thus, the profile log pseudo-likelihood function is concave (and strictly concave under mild conditions), so the maximum pseudo-likelihood estimate of $\psi$ (provided it exists) can be found by a numerical optimization method where evaluating \eqref{e:pseudo} and \eqref{e:score} involves approximating the integrals by numerical methods \citep{PL2}. 

Second, we condition on both $X_{\partial W_{\ominus R}}=x_{\partial W_{\ominus R}}$ and $N(W_{\ominus R})=m$. For $i=1,\ldots,m$, define $X_{-i}\coloneqq (X_1,\ldots,X_{i-1},X_{i+1},\ldots,X_m)$ and $x_{-i}\coloneqq (x_1,\ldots,x_{i-1},x_{i+1},\ldots,x_m)$. The random vectors $X_{-1},\ldots,X_{-m}$ are identically distributed. Further, $X_i$ conditioned on both $X_{\partial W_{\ominus R}}=x_{\partial W_{\ominus R}}$, $N(W_{\ominus R})=m$ and $X_{-i}=x_{-i}$ depends only on $x\setminus\{x_i\}$ and has probability density function 
\begin{equation*}\label{e:f_psi}
f_{\psi}(u\,|\,x\setminus\{x_i\})\coloneqq a(x\setminus\{x_i\},u)\exp(\psi^Ts(x\setminus\{x_i\},u))/c_{\psi}(x\setminus\{x_i\})
\end{equation*}
for $u\in W_{\ominus R}$ where 
\[
c_{\psi}(x\setminus\{x_i\})=\int_{W_{\ominus R}}a(x\setminus\{x_i\},u)\exp(\psi^Ts(x\setminus\{x_i\},u))\,\mathrm du. 
\]
Now, the logarithm of Besag's pseudo-likelihood function \citep{Besag} as defined by the product of the `full conditionals' $f_{\psi}(u\,|\,x\setminus\{x_i\})$ ($i=1,\ldots,m$) becomes 
\begin{equation}\label{e:besag}
pl_m(\psi)\coloneqq  \psi^T \sum_{i=1}^m s(x\setminus\{x_i\},x_i)-\sum_{i=1}^m\log c_{\psi}(x\setminus\{x_i\})
\end{equation}
when we omit the term $\sum_{i=1}^m\log a(x\setminus\{x_i\},x_i)$, which only depends on the data. Assuming we can interchange differentiation and integration, the pseudo-score corresponding to \eqref{e:besag} is
\begin{align*}\label{e:pseudoscore}
s_m(\psi)\coloneqq \frac{\partial}{\partial\psi}pl_m(\psi)=\sum_{i=1}^m s(x\setminus\{x_i\},x_i)-
\sum_{i=1}^m \int_{W_{\ominus R}}s(x\setminus\{x_i\},u)f_{\psi}(u\,|\,x\setminus\{x_i\})\,\mathrm du,
\end{align*}
and its derivative is
\begin{align*}
\frac{\partial}{\partial\psi^T} s_m(\psi)&=-\sum_{i=1}^m \Bigg (\int_{W_{\ominus R}}s(x\setminus\{x_i\},u)s(x\setminus\{x_i\},u)^Tf_{\psi}(u\,|\,x\setminus\{x_i\})\,\mathrm du\\
&\quad - \int_{W_{\ominus R}}s(x\setminus\{x_i\},u)f_{\psi}(u\,|\,x\setminus\{x_i\})\,\mathrm du\int_{W_{\ominus R}}s(x\setminus\{x_i\},u)^Tf_{\psi}(u\,|\,x\setminus\{x_i\})\,\mathrm du\Bigg)\\
&=-\sum_{i=1}^m {\mathrm{Var}}_\psi \left[s(x\setminus\{x_i\},X_i)\,|\,x\setminus\{x_i\}\right].
\end{align*}
So, the log pseudo-likelihood is again concave (and strictly concave under mild conditions) and can be optimized numerically using numerical evaluation of integrals, but the computations may be more cumbersome compared to \eqref{e:pseudo} and \eqref{e:score} because we need to evaluate more integrals. However, it is still easier than using maximum likelihood estimation.

\end{document}